\documentclass{article}
\usepackage{apacite}
\usepackage{amsmath}
\usepackage[margin=0.75in]{geometry}
\usepackage{graphicx}
\usepackage{newtxtext,newtxmath}
\usepackage{titlesec}
\usepackage[export]{adjustbox}
\usepackage{natbib}
\usepackage[title]{appendix}
\usepackage{setspace}
\usepackage{hyperref}
\usepackage{multirow}
\usepackage{caption}
\begin{document}

\title{A model of the Unity High Definition Render Pipeline, with applications to flat-panel and head-mounted display characterization}
\author{Richard F. Murray \\
Department of Psychology and Centre for Vision Research, York University \\
rfm@yorku.ca}
\date{}
\maketitle

\begin{center} \textbf{Abstract} \end{center}
\begin{quote}
Game engines such as Unity and Unreal Engine have become popular tools for creating perceptual and behavioral experiments in complex, interactive environments. They are often used with flat-panel displays, and also with head-mounted displays. Here I describe and test a mathematical model of luminance and color in Unity's High Definition Render Pipeline (HDRP). I show that the HDRP has several non-obvious features, such as nonlinearities applied to material properties and rendered values, that must be taken into account in order to show well-controlled stimuli. I also show how the HDRP can be configured to display gamma-corrected luminance and color, and I provide software to create the specialized files needed for gamma correction.
\end{quote}
\vspace{6pt}

%
%

\section{Introduction} \label{sec:introduction}

Vision science relies on a wide range of display technologies, such as stereoscopes, Maxwellian-view systems, and computer-controlled flat-panel displays, all of which have played a central role in developing and testing current theories of vision. In recent years, programming tools for creating interactive environments, such as Unity and Unreal Engine, have been widely used for behavioral and perceptual experiments \citep{Unity2022p3, Unreal2024}. These frameworks were originally developed for gaming applications, but they have been useful for investigating abilities that support interactions with  complex scenes, such as navigation and color constancy. These tools have been used to create experiments with traditional flat-panel displays, and also with newer display technologies such as virtual reality (VR) and augmented reality (AR) \citep{kimura2017, gilrodriguez2022, creem2022, rzepka2023, patel2024}.

One challenge with using these programming environments for research is showing precisely controlled stimuli. Game engines, as well as consumer VR and AR devices, have been primarily developed for entertainment, so in research applications they often require specialized methods for precise control of stimuli. In previous work, colleagues and I described methods for controlling luminance in Unity's Built-in Render Pipeline (BRP; Murray, Patel, \& Wiedenmann, \citeyear{murray2022}). The BRP is an older pipeline, and much current development work in Unity focuses on other pipelines, such as the High-Definition Render Pipeline (HDRP), which aims to be a physics-based renderer, and includes features such as raytracing that are not present in the BRP. The HDRP also has the advantage of using meaningful photometric units for some scene properties such as lighting intensity.

Here I describe and test a mathematical model of rendering in the HDRP, with the goal of providing a model and related software tools that researchers can use to create experiments with precise control of luminance and color. I show that the HDRP has several non-obvious features, such as nonlinearities applied to material properties and rendered values, that make it important to have a model of rendering in order to maintain control over stimuli. I show how the rendering pipeline can be configured to show stimuli with gamma-corrected luminance and color. I also show how to test a wide range of hypotheses about the HDRP, so that researchers can examine additional features that are relevant to their own work. Supporting Information with Unity projects and analysis code is available at \href{https://doi.org/10.17605/OSF.IO/NHPUA}{https://doi.org/10.17605/OSF.IO/NHPUA}.

%
%

\section{Previous work}

\cite{brainard2002} is a classic reference on display characterization for perceptual experiments. It focuses on characterization of cathode-ray tube (CRT) displays, but the modelling framework that it describes, as well as its discussion of issues such as spatial interactions between neighboring pixels, are highly relevant to more recent modalities as well. This paper is a valuable starting point for learning about display characterization.

\cite{clausen2019} tested a mathematical display model similar to the one described by \cite{brainard2002}, for two models of head-mounted VR displays, and found that the model accurately predicted displayed colors. However, they also found that it was important to set model parameters using characterization measurements for individual displays, instead of relying on standards such as sRGB or Adobe RGB, as standard models led to easily perceived errors in displayed color. \cite{DiazBarrancas2024} extended this work, making thorough color characterization measurements of three head-mounted display models, and examining display features such as primary spectra, color gamut, and spatial uniformity.

\cite{toscani2019} made photometric and spectral measurements from a VR headset controlled by Unreal Engine 4. They found that with Unreal's default settings, luminance was a nonlinear and non-additive function of the simulated albedo in rendered scenes, making it difficult to generate stimuli with desired colors. However, with some of Unreal's post-processing operations disabled, luminance was a linear and approximately additive function of simulated albedo, and displayed colors were more easily controlled. \cite{gilrodriguez2022} built on this work, showing that colors generated by Unreal in a complex virtual environment could be predicted from rendering parameters, and that observers were able to achieve naturalistic levels of color constancy in a color matching experiment in VR.

\cite{zaman2023} took an alternative approach to creating calibrated stimuli in Unreal, and showed how specialized virtual materials could be used to control the chromaticity of selected objects separately from the rest of a virtual scene, so that post-processing operations did not have to be disabled for the whole scene. \cite{duay2025} adapted this approach to Three.js, a web-based render engine, and showed how it could be used with standard materials instead of more exotic `unlit' material types.

\cite{DiazBarrancas2020, DiazBarrancas2021} showed how hyperspectral rendering methods could be implemented in Unity, using scripts that mapped hyperspectral lighting and surface information to sRGB color coordinates in real time. \cite{kim2021} evaluated several material and lighting settings in Unity, and measured their effects on the chromaticity of rendered objects, with the goal of finding settings that led to well-controlled and predictable colors in renderings of uniform patches and biological tissue samples.
 
\cite{murray2022} developed methods for controlling luminance in Unity's Built-in Render Pipeline. We described procedures for making luminance characterization measurements, and showed how to use Unity's `color grading' feature to make displayed luminance proportional to rendered values. The present paper extends that work, broadening its methods to handle both luminance and color, and using the newer HDRP instead of the BRP.

%
%

\section{A model of the HDRP}

In its default configuration, an HDRP project has features that make it difficult to control luminance and color. For example, a dynamic exposure feature automatically adjusts the gain on rendered images, so that as a user moves through scenes with varying levels of simulated lighting, images are mapped to a displayable range on the monitor. This is useful for applications such as gaming, but it removes the control we often need in perceptual experiments. In Appendix \ref{appendix:configure}, I describe configuration settings for HDRP projects that make it easier to control luminance and color. The model of the HDRP that I present in the following sections assumes that these settings have been made. Throughout I use Unity 2022.3.10, which includes HDRP 14.0.9.

In the following sections I outline a model of the HDRP for some common lighting conditions and materials. To anticipate, a key part of the model is a description of Unity's \textit{tonemapping} mechanism, which allows the user to apply a configurable pointwise nonlinearity to the rendered image, and provides a method of implementing gamma correction.

%
%

\subsection{Lighting and material model}

\underline{Lambertian material.} In order to generate precisely controlled stimuli with the HDRP, we need a model of how it uses virtual lights and surfaces to compute red, green, and blue values at each pixel. Here I start by examining how the HDRP renders Lambertian surfaces.

In a typical Lambertian shading model, a material has an albedo $m$ that indicates the proportion of light it reflects in the visible wavelength range. The material surface also has a unit normal vector $\mathbf{n}$ at each point. In a simple lighting model, lighting consists of a directional source with maximum illuminance $d$ in direction given by unit vector $\mathbf{l}$, and an ambient source with illuminance $a$ in all directions. According to the Lambertian model, the luminance of the surface from any viewing direction is
\begin{equation}
u = \frac{m}{\pi} ( d \max( \cos \theta, 0) +a )
\label{eqn:lambert1}
\end{equation}
where $\theta$ is the angle between the lighting direction $\mathbf{l}$ and the surface normal $\mathbf{n}$ (i.e., $\cos \theta = \mathbf{l} \cdot \mathbf{n}$), and $\max(x,y)$ is the greater of $x$ and $y$ \citep{McCluney2014, pharr2023}. Here the factor $\cos \theta$ models the falloff of illumination received from a point light source as a function of its angle $\theta$ relative to the surface normal $\mathbf{n}$, and the $\max$ function incorporates the fact that a surface receives no illumination from a light source at an angle greater than 90$^\circ$ relative to the normal, i.e., behind the surface.

The HDRP supports Lambertian shading, but it parameterizes materials and lights differently than in equation (\ref{eqn:lambert1}). Unity has a material type called HDRP/ArnoldStandardSurface/ArnoldStandardSurface, which is Lambertian with its default parameter settings. This material can be assigned a triplet $\mathbf{m} = (m_r, m_g, m_b)$ of albedo-like components for the three color channels, where each component ranges over the unit interval [0, 1]. Each point on a surface also has a unit normal vector $\mathbf{n}$, usually computed automatically from the shape that the material is assigned to. A directional light source has an intensity\footnote{`Intensity' is the term used in the Unity interface for a multiplicative factor in light source magnitude, so I will use that term as well. Photometrically, it is something quite different, namely luminous power per unit of solid angle. I point this out because the HDRP sometimes uses photometrically correct units, but not in this case.} $i_d$ and color components $\mathbf{d} = (d_r, d_g, d_b)$. The direction of the light is controlled by rotations around the $x$, $y$, and $z$ axes, which result in a lighting direction $\mathbf{l}$. An ambient source has intensity $i_a$ and color components $\mathbf{a} = (a_r, a_g, a_b)$. (In Appendix \ref{appendix:configure}, I describe how to create these materials and light sources.) With these parameters, the rendered value in each color channel $k$ is
\begin{equation}
u_k = \frac{c \: s(m_k) ( i_d \, s(d_k) \max( \cos \theta, 0 ) / \pi + i_a a_k)}{2^e}
\label{eqn:lambert2}
\end{equation}
Here the index $k$ stands for the three color channel indices $r$, $g$, and $b$. I call $u_k$ the `unprocessed' color coordinates, because as we will see, this stage of rendering is followed by post-processing operations.

There are several important differences between equations (\ref{eqn:lambert1}) and (\ref{eqn:lambert2}). First, equation (\ref{eqn:lambert2}) has a scale factor $c = 0.822$, which I determined empirically (see Section \ref{section:tests}) and whose origin is unclear. It may be related to properties of the virtual camera.

Second, the albedo-like components $m_k$ and the directional light color components $d_k$ are transformed by a nonlinearity $s$, which is the `transfer function' of the sRGB color space \citep{anderson1996}. This nonlinearity consists of a small linear segment followed by a power function with exponent 2.4. It can be approximated by $s(x) \simeq x^{2.2}$. This transformation is why I call the components $m_k$ albedo-\underline{like}, and clearly it is important to take this nonlinearity into account when creating stimuli. For example, a value $m_k = 0.5$ results in an albedo of just $s(0.5) = 0.21$. Note that the ambient lighting components $a_k$ are not transformed this way. Furthermore, the directional light components $d_k$ range over the unit interval [0, 1], whereas the ambient components $a_k$ range over [0, $\infty$).
 
Third, the directional and ambient lights differ by a factor of $\pi$ in how they contribute to the unprocessed value $u_k$. One possible reason for this is that a uniform ambient light with visible luminance $\ell$ in all directions produces an incident illuminance $\pi \ell$ \citep{McCluney2014}. Thus the product $i_d s(d_k)$ in equation (\ref{eqn:lambert2}) can be seen as representing the maximum \textit{incident illuminance} of the directional light, which occurs for a surface whose normal is in the lighting direction, $\mathbf{n} = \mathbf{l}$, whereas the product $i_a a_k$ represents the \textit{visible luminance} of the ambient light in all directions, rather than the incident illuminance it creates.

Fourth, equation (\ref{eqn:lambert2}) includes a scale factor $2^e$, where $e$ is the exposure setting described in Appendix \ref{appendix:configure}.

Table \ref{table:variables} lists the variables of this model and the range of values they take, as well as additional variables introduced in later sections. I arrived at equation (\ref{eqn:lambert2}) empirically, by trial and error. In Section \ref{section:tests}, I report tests that largely validate equation (\ref{eqn:lambert2}) as a model of Lambertian rendering in the HDRP, although I note some small departures as well.

\begin{table}[t]
\caption{Variables in the HDRP model}
\begin{tabular}{ c l l l }
\hline \hline
& \textbf{variable} & \textbf{value range} & \textbf{description} \\
 \hline
& $\mathbf{m} = (m_r, m_g, m_b)$ & $[0, 1]^3$ & material color \\  
& $\mathbf{n} = (n_x, n_y, n_z)$ & unit sphere & surface normal \\
& $\mathbf{d} = (d_r, d_g, d_b)$ & $[0, 1]^3$ & directional light color \\
& $i_d$ & $[0, \infty)$ & directional light intensity \\
& $\mathbf{l} = (l_x, l_y, l_z)$ & unit sphere & directional light direction \\
& $\mathbf{a} = (a_r, a_g, a_b)$ & $[0, \infty)^3$ & ambient light color \\
& $i_a$ & $[0, \infty)$ & ambient light intensity \\
& $c$ & 0.822 & rendering constant \\
& $s$ & $[0, 1] \rightarrow [0, 1]$ & sRGB transfer function \\
& $\mathbf{s}$ & $[0, 1]^3 \rightarrow [0, 1]^3$ & sRGB transfer function, vectorized \\
& $e$ & $(-\infty, \infty)$ & exposure \\
\multirow{-12}{*}{\rotatebox[origin=c]{90}{\textbf{scene elements}}} & $\mathbf{u} = (u_r, u_g, u_b)$ & $[0, \infty)^3$ & unprocessed color \\
\hline
& $\mathbf{t} = (t_r, t_g, t_b)$ & $[0, \infty)^3$ & tonemapped color \\
& $\mathbf{f}$ & $[0, \infty)^3 \rightarrow [0, 1]^3$ & tonemapping function \\
& $f$ & $[0, \infty) \rightarrow [0, 1]$ & tonemapping function, single component \\
& $u^*_i$ & $[0, \infty)$ & tonemapping knot points, single component \\
& $\mathbf{u}_{ijk} = (u^r_{ijk}, u^g_{ijk}, u^b_{ijk})$ & $[0, \infty)^3$ & tonemapping knot points \\
& $\mathbf{t}_{ijk} = (t^r_{ijk}, t^g_{ijk}, t^b_{ijk})$ & $[0, 1]^3$ & tonemapping output at knot points \\
\multirow{-7}{*}{\rotatebox[origin=c]{90}{\textbf{post-processing}}} & $\mathbf{v} = (v_r, v_g, v_b)$ & $[0, 1]^3$ & post-processed color \\
\hline
& $h$ & $[0, 1] \rightarrow [0, 1]$ & achromatic activation function \\
& $L_0$ & $[0, \infty)$ & minimum displayable luminance \\
& $L_1$ & $[0, \infty)$ & luminance range (maximum minus minimum) \\
& $L$ & $[0, \infty)$ & displayed luminance \\
& $h_r, h_g, h_b$ & $[0, 1] \rightarrow [0, 1]$ & chromatic activation functions \\
& $\mathbf{p} = (p_r, p_g, p_b)$ & $[0, 1]^3$ & chromatic channel activations \\
& $\mathbf{r}$, $\mathbf{g}$, $\mathbf{b}$, $\mathbf{z}$ & $[0, \infty)^3$ & display primaries and background term (CIE XYZ) \\
\multirow{-8}{*}{\rotatebox[origin=c]{90}{\textbf{display}}} & $\mathbf{x}$ & $[0, \infty)^3$ & displayed color coordinates (CIE XYZ) \\
\hline \hline
\end{tabular}
\label{table:variables}
\end{table}

\underline{Unlit material.} The HDRP also has a material type called HDRP/Unlit, which is not affected by lighting or by the exposure setting. Its rendering model is very simple, which makes it useful for testing hypotheses about the render pipeline. An unlit material is assigned a color triplet $\mathbf{m} = (m_r, m_g, m_b)$, and its unprocessed value in each color channel is
\begin{equation}
u_k = s( m_k )
\label{eqn:unlit}
\end{equation}
Here $m_k$ and $u_k$ both range over [0, 1], and $s$ is the sRGB transfer function introduced above. In Section \ref{section:tests} I report tests that validate this model as well, again with some small departures.

%
%

\subsection{Post-processing model}

The unprocessed triplets $\mathbf{u} = (u_r, u_g, u_b)$ in equations (\ref{eqn:lambert2}) and (\ref{eqn:unlit}) are transformed into post-processed triplets $\mathbf{v} = (v_r, v_g, v_b)$ as follows.
\begin{equation}
\mathbf{v} = \mathbf{s}^{-1} ( \mathbf{f}( \mathbf{u} ) )
\label{eqn:post1}
\end{equation}
Here $\mathbf{s}^{-1}$ is the inverse of a vectorized version $\mathbf{s}$ of the sRGB transfer function $s$, that applies $s$ to each component: $\mathbf{s}((x, y, z)) = (s(x), s(y), s(z))$. The function $\mathbf{f}$ is a \textit{tonemapping} function that transforms unprocessed color triplets, and is chosen in the tonemapping override described in Appendix \ref{appendix:configure}. The HDRP provides a few standard choices for the tonemapping function, and also allows the user to construct a custom tonemapping function using a `cube' file. Below I show how to use tonemapping via a cube file to implement gamma correction.

The unprocessed components $u_k$ range over $[0, \infty)$, whereas $v_k$ range over [0, 1], so one role of this post-processing stage is to map the unprocessed values to a limited, displayable range.

In this first pass at describing the HDRP, I assume that we are interested in achromatic stimuli where $u_r = u_g = u_b$, and that we wish to gamma-correct the luminance of displayed stimuli. In Section \ref{section:chromatic} I generalize this approach to chromatic stimuli and colorimetric gamma correction. As part of this simplification, I assume that we choose a tonemapping function that applies the same nonlinearity $f$ to all three color channels: $\mathbf{f}(\mathbf{u}) = (f(u_r), f(u_g), f(u_b))$. Equation (\ref{eqn:post1}) then becomes
\begin{equation}
v_k = s^{-1} ( f( u_k ) )
\label{eqn:post2}
\end{equation}
In this achromatic model, I assume that $u_r = u_g = u_b$, and so $v_r = v_g = v_b$ as well.

One fact that I have not incorporated into the model is that in Unity's default configuration, the post-processed values $v_k$ are quantized to multiples of 1/255. This point is relevant to the design of some experiments where observers make fine discriminations.

%
%

\subsection{Display model}

The post-processed triplets $\mathbf{v}$ are written to a buffer and shown on a display, such as a flat-panel LCD monitor or VR headset. The display transforms the triplets $\mathbf{v}$ into physical luminances and chromaticities, so we also need a model of the display. The XYZ color coordinates $\mathbf{x}$ of the displayed stimulus are some function of $\mathbf{v}$. Continuing for now to model achromatic stimuli, the luminance $L$ of the displayed stimulus can be written as the following function the post-processed components.
\begin{equation}
L = L_1 h( v_k ) + L_0
\label{eqn:display}
\end{equation}
Here $L_0$ is the minimum displayable luminance and $L_1$ is the range of displayable luminances, i.e., the maximum minus the minimum. I will call $h$ the `activation function' of the display. It rises from $h(0) = 0$ to $h(1) = 1$. It is often approximately given by $h(x) = x^{2.2}$, but it is more reliable to measure the activation function as part of gamma correction, as I describe below.

%
%

\subsection{Summary}

To summarize this model of the HDRP, we start with models for Lambertian and unlit materials that map lighting and material parameters to unprocessed color coordinates.
\begin{equation}
u_k = \frac{c \: s(m_k) ( i_d \, s(d_k) \max( \cos \theta, 0 ) / \pi + i_a a_k)}{2^e}
\end{equation}
\begin{equation}
u_k = s( m_k )
\end{equation}
This is followed by post-processing, which includes tonemapping $f$ that can be configured by the user, and a fixed nonlinearity $s^{-1}$.
\begin{equation}
v_k = s^{-1} ( f( u_k ) )
\end{equation}
Finally, the post-processed values drive a display, which results in a physical luminance.
\begin{equation}
L = L_1 h( v_k ) + L_0
\end{equation}
Table \ref{table:variables} lists the variables of this model.

%
%

\section{Tonemapping} \label{section:tonemapping}

The HDRP's tonemapping interface, described in Appendix \ref{appendix:configure}, allows the user to choose from a number of `modes' that transform unprocessed values $\mathbf{u}$ as in equation (\ref{eqn:post1}). One option is `None', which disables tonemapping and sets $\mathbf{f}(\mathbf{u}) = \mathbf{u}$. Another option is `External', which applies the transformation specified in a user-supplied text file that follows the `cube' format \citep{AdobeCube2013}.

External tonemapping works as follows. The HDRP has a fixed list of floating point values $u^*_i$, where $i$ ranges from 1 to $n$. (The default value is $n = 32$, which is the setting I will assume here. The value can be adjusted in the project settings.) From this list it constructs a 3D grid of unprocessed color triplets $\mathbf{u}_{ijk} = (u^r_{ijk}, u^g_{ijk}, u^b_{ijk} ) = (u^*_i, u^*_j, u^*_k)$, where each subscript ranges from 1 to $n$. A user-supplied cube file specifies an equally-sized 3D grid of replacement color triplets $\mathbf{t}_{ijk} = (t^r_{ijk}, t^g_{ijk}, t^b_{ijk})$. Tonemapping maps elements of the first grid to corresponding elements of the second grid: $\mathbf{f}(\mathbf{u}_{ijk}) = \mathbf{t}_{ijk}$. Unprocessed color triplets $\mathbf{u}$ that are not equal to one of the triplets $\mathbf{u}_{ijk}$ are mapped to outputs by linear interpolation from nearby input points $\mathbf{u}_{ijk}$ and output points $\mathbf{t}_{ijk}$. In interpolation frameworks such as this one, the points $\mathbf{u}_{ijk}$ are often called `knot points'.

In Section \ref{section:tests}, I report tests that find the knot points points $\mathbf{u}_{ijk}$ used by the HDRP, and validate this model of tonemapping in External mode. Tonemapping can be used for gamma correction, as described in the following section.

%
%

\section{Achromatic gamma correction}

For achromatic stimuli, the goal of gamma correction is to make the physical luminance $L$ of each pixel proportional to the unprocessed value $u_k$ that is rendered from scene parameters such as material properties and lighting. Combining the post-processing and display models in equations (\ref{eqn:post2}) and (\ref{eqn:display}), the relationship between $L$ and $u_k$ is
\begin{equation}
L = L_1 h( s^{-1} ( f( u_k ) ) ) + L_0
\label{eqn:gamma1}
\end{equation}
Alternatively, we can define $w = L_0/L_1$, and write equation (\ref{eqn:gamma1}) as
\begin{equation}
L = L_1 ( h( s^{-1} ( f( u_k ) ) ) +w)
\label{eqn:gamma2}
\end{equation}
This provides a solution for gamma correction: we can make $L$ proportional to $u_k$ by setting the tonemapping function $f$ to
\begin{equation}
f(u_k) = s(h^{-1}( (1+w)\frac{u_k}{r} - w))
\label{eqn:tonemap1}
\end{equation}
where $r$ is a constant. Then equation (\ref{eqn:gamma2}) reduces to
\begin{equation}
L = (L_0 + L_1) \frac{u_k}{r}
\label{eqn:proportional1}
\end{equation}
and $L$ is proportional to $u_k$.

One point I have glossed over in the previous paragraph is that $h^{-1}$ maps [0, 1] to [0, 1], but for small values of $u_k$, the input to $h^{-1}$ in equation (\ref{eqn:tonemap1}) is negative. (Consider $u_k = 0$, for instance.) This occurs for small values of $u_k$ that would ideally be mapped to luminances less than $L_0$, the minimum displayable luminance. A natural solution is to revise equation (\ref{eqn:tonemap1}) so that the input to $h^{-1}$ is clipped to non-negative values, as follows.
\begin{equation}
f(u_k) = s(h^{-1}( \max( (1+w)\frac{u_k}{r} - w, 0)))
\label{eqn:tonemap2}
\end{equation}
Then, as $u_k$ approaches zero, the luminance goes no lower than $L_0$, the lowest displayable luminance, as we might expect. Equation (\ref{eqn:proportional1}) is then replaced by
\begin{equation}
L = \begin{cases}
\hspace{0.2cm} (L_0 + L_1) \frac{u_k}{r} & u_k \geq u_0 \equiv \frac{rw}{1+w}  \\
\hspace{0.2cm} L_0 & u_k < u_0
\end{cases}
\label{eqn:proportional2}
\end{equation}
and $L$ is proportional to $u_k$, except when $u_k < u_0$, in which case $L = L_0$. I will consider typical values for the cutoff point $u_0$ below, but to anticipate, it is typically very small compared to the displayable range of $u_k$.

To create the tonemapping function defined in equation (\ref{eqn:tonemap2}) we need to know $s$, $h$, $L_0$, $L_1$, and $r$. The function $s$ is fixed, and tests in Section \ref{section:tests} show that it is the sRGB transfer function. We can find $h$, $L_0$, and $L_1$ by measuring the mapping from post-processed values $v_k$ to luminance $L$ on the display. To make these measurements, we can display post-processed values $v_k$ using the unlit material type discussed earlier. Equations (\ref{eqn:unlit}) and (\ref{eqn:post2}) show that if tonemapping is disabled (which sets $f(x) = x$), then the post-processed values of an unlit material with color coordinates $m_k$ are
\begin{equation}
v_k = s^{-1}(f(s(m_k))) = m_k
\label{eqn:unlit2}
\end{equation}
That is, assigning an unlit material color coordinates $m_k$ results in post-processed values $v_k = m_k$. To measure $h$, $L_0$, and $L_1$, we can display surfaces with a range of post-processed values $v_k$, and measure the resulting luminances.

Finally, the constant $r$ can chosen arbitrarily, and it determines the range of unprocessed values $u_k$ that is mapped to the displayable range of luminances. Equation (\ref{eqn:proportional2}) shows that the maximum luminance $L_0 + L_1$ is shown when $u_k / r = 1$, so the displayable range of $u_k$ is $[0, r]$. We can choose $r=1$, in which case the displayable range of $u_k$ is [0, 1].

Now we can consider typical values for the cutoff point $u_0$ in equation (\ref{eqn:proportional2}). We can choose $r = 1$, and typically $w = L_0/L_1$ is no larger than 0.01, which results in $u_0 \approx 0.01$. In this case luminance $L$ equals $L_0$ for $u_k$ in [0, 0.01], and $L$ is proportional to $u_k$ over the much larger range [0.01, 1].

By choosing appropriate output triplets $\mathbf{t}_{ijk}$ for tonemapping in External mode, we can create an approximation to the tonemapping function defined for gamma correction in equation (\ref{eqn:tonemap2}). The most straightforward approach is for the tonemapping function to process each color channel independently, and to map the input points $\mathbf{u}_{ijk} = (u^*_i, u^*_j, u^*_k)$ to the values defined by equation (\ref{eqn:tonemap2}):
\begin{equation}
\mathbf{t}_{ijk} = (t^r_{ijk}, t^g_{ijk}, t^b_{ijk}) = ( f(u^*_i), f(u^*_j), f(u^*_k) )
\label{eqn:tonemap_point}
\end{equation}
This gives an approximation to the tonemapping function in equation (\ref{eqn:tonemap2}). We can improve this approximation slightly by taking the $\mathbf{t}_{ijk}$ in equation (\ref{eqn:tonemap_point}) as initial values in an optimization problem that finds the $\mathbf{t}_{ijk}$ that minimize the difference between the approximation and the function defined by equation (\ref{eqn:tonemap2}), using a global error measure such as the sum-of-squares difference between the functions. In Section \ref{section:tests_gamma}, I show results from a Unity project included in the Supporting Information that takes this approach to gamma correction.

%
%

\section{A chromatic model} \label{section:chromatic}

\begin{figure*}[h!]
\centering
\includegraphics[width=18cm, left]{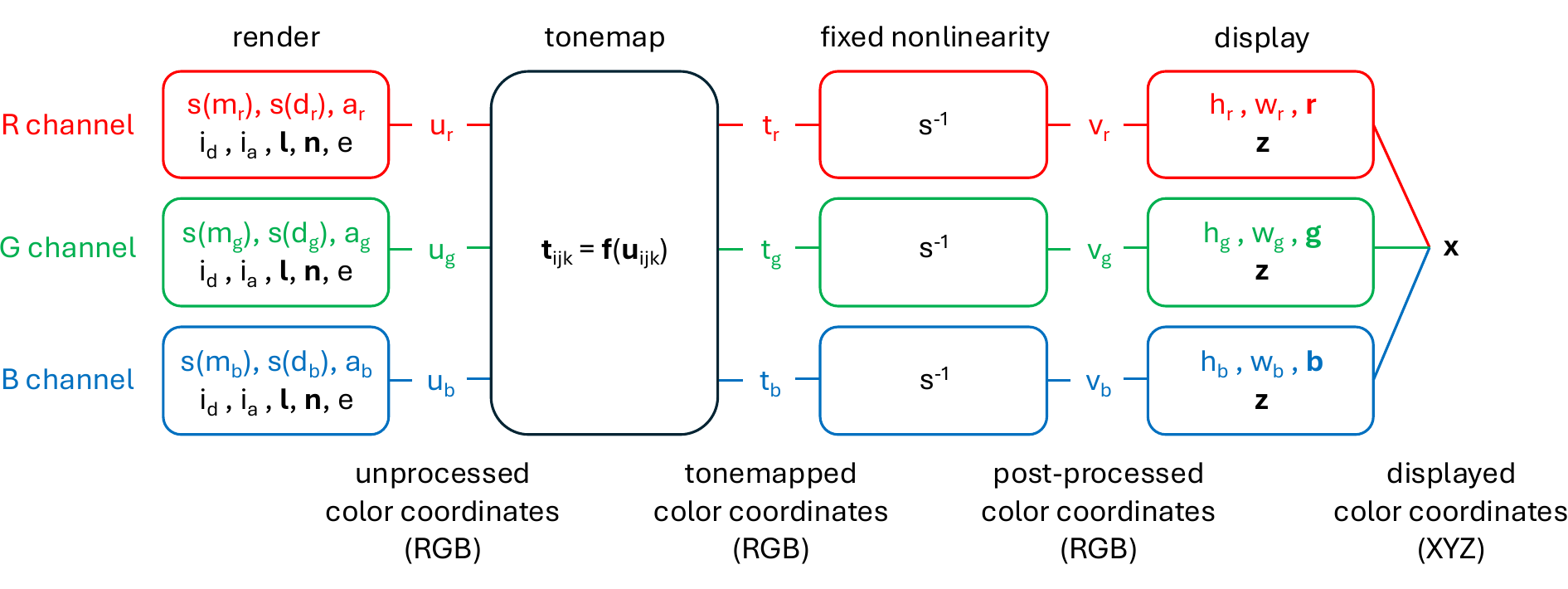}
\caption{A visual representation of the chromatic HDRP model. Red, green, and blue represent elements that are specific to individual chromatic channels, and black represents elements that are common to all three channels. Each box shows the variables that affect its computation. The variables are summarized in Table \ref{table:variables}.}
\label{figure:chromatic_model}
\end{figure*}

Here I extend the model in the previous sections to include color and methods for chromatic gamma correction. Figure \ref{figure:chromatic_model} gives a graphical overview of the chromatic HDRP model.

\subsection{A chromatic display model}

The CIE XYZ chromaticity coordinates $\mathbf{x}$ of a stimulus are some function of the post-processed color coordinates $\mathbf{v}$. In many displays, each coordinate $v_k$ controls the intensity of a primary color ($\mathbf{r}$, $\mathbf{g}$, or $\mathbf{b}$), the three scaled primaries sum, and there is also a small, fixed background term ($\mathbf{z}$).
\begin{equation}
\mathbf{x} = h_r(v_r) \, \mathbf{r} + h_g(v_g) \, \mathbf{g} + h_b(v_b) \, \mathbf{b} \: + \: \mathbf{z}
\label{eqn:chromatic_model1}
\end{equation}
The activation functions $h_k$ describe how the primary intensities depend on $v_k$. We can also refer to the `activation' $p_k = h_k(v_k)$ of each chromatic channel. Equation (\ref{eqn:chromatic_model1}) expresses the important assumptions that (a) the chromaticities of the primaries do not change with their level of activation, and (b) the three chromatic channels combine in a purely additive fashion.

If the background term can be expressed as a sum of the primaries, $\mathbf{z} = w_r \, \mathbf{r} + w_g \, \mathbf{g} + w_b \, \mathbf{b}$, then it can be absorbed into the primary coefficients.
\begin{equation}
\mathbf{x} = (h_r(v_r)+w_r) \, \mathbf{r} + (h_g(v_g)+w_g) \, \mathbf{g} + (h_b(v_b)+w_b) \, \mathbf{b}
\label{eqn:chromatic_model2}
\end{equation}
The three components of this equation consist of an activation term and a fixed background term, much like equation (\ref{eqn:display}) in the achromatic luminance model. This suggests that we can carry out chromatic gamma correction simply by applying the method described for achromatic gamma correction to each color channel independently, and this is the approach I will take.

\subsection{Chromatic gamma correction}

One possible goal for chromatic gamma correction is that the primary coefficients in equation (\ref{eqn:chromatic_model2}) should be proportional to the unprocessed color coordinates $u_k$.
\begin{equation}
h_k(v_k)+w_k \: \propto \: u_k
\label{eqn:chromatic_goal1}
\end{equation}
We can also choose the constant of proportionality in equation (\ref{eqn:chromatic_goal1}) so that at $u_k = 1$, the corresponding primary is at its maximum intensity, e.g., $u_r = 1$ displays the color $(1+w_r)\mathbf{r}$.

Equation (\ref{eqn:post1}) describes how post-processing maps unprocessed color coordinates $\mathbf{u}$ to post-processed coordinates $\mathbf{v}$. I repeat it here:
\begin{equation}
\mathbf{v} = \mathbf{s}^{-1} ( \mathbf{f}( \mathbf{u} ) )
\end{equation}
Our goal is to choose the tonemapping function $\mathbf{f}$ so that $u_k$ and $v_k$ are related as in equation (\ref{eqn:chromatic_goal1}). We can choose a tonemapping function that consists of three scalar functions $f_k$ that are applied to the three color channels independently.
\begin{equation}
v_k = s^{-1}(f_k(u_k))
\end{equation}
Following the approach used for achromatic gamma correction, I define the components of the tonemapping function to be
\begin{equation}
f_k(u_k) = s(h_k^{-1}( \max( (1+w_k) u_k - w_k, 0)))
\label{eqn:tonemap3}
\end{equation}
With this definition of $f_k$,
\begin{align*}
h_k( v_k ) & = h_k( s^{-1}(f_k(u_k)) ) \\
& = \begin{cases}
\hspace{0.2cm} (1+w_k) u_k - w_k & u_k \geq w_k / (1+w_k) \\
\hspace{0.2cm} 0 & \text{otherwise}
\end{cases}
\end{align*}
and $h_k(v_k) + w_k$ is proportional to $u_k$, as was the goal in equation (\ref{eqn:chromatic_goal1}), except at small values of $u_k$. In the latter case, the displayed stimulus $\mathbf{x}$ cannot approach zero, and is limited by the background component $\mathbf{z}$. Furthermore, when $u_k = 1$, the corresponding primary is displayed with maximum intensity $h_k(v_k) + w_k = 1 + w_k$, as was also the goal.

Additional goals are possible for chromatic gamma correction. We might also want to control the chromaticity displayed when $u_r = u_g = u_b$ (i.e., the white point), and to use arbitrary color primaries for $\mathbf{r}$, $\mathbf{g}$, and $\mathbf{b}$ instead of the physical primaries of the display. The mapping that can be specified in External tonemapping mode seems to be flexible enough to incorporate these features, but I will not develop them here.

%
%

\section{Tests of the HDRP model} \label{section:tests}

Here I report tests of the model described in the previous sections. The Supporting Information includes all Unity projects and analysis scripts required to carry out these tests, as well as detailed instructions.

%
%

\subsection{Tests without tonemapping} \label{section:tests_without_tonemapping}

To test the HDRP model, I created a Unity project (render\_random in the Supporting Information) where the scene consists of a plane under directional and ambient lights. The project assigns random color coordinates $m_k$ to the plane, and rotates it to a random orientation $\mathbf{n}$. It also assigns random color coordinates $d_k$, $a_k$ and intensities $i_d$, $i_a$ to the directional and ambient lights, and a random direction $\mathbf{l}$ to the directional light. It then captures the post-processed color coordinates $v_k$ at one pixel in the rendered framebuffer image, using the Unity function Texture2D.ReadPixels(). It records the random scene parameters and the rendered framebuffer values in a text file. This process is repeated many times, providing many samples of randomly generated material and lighting parameters, and resulting rendered values. Thus the values that are directly available for testing the HDRP model are the parameters for lighting, materials, etc. that are assigned to the scene, and the post-processed values $v_k$ read from the framebuffer. The model can be tested by predicting the post-processed values $v_k$ from the scene parameters, and also by testing for consistency when predicting internal values such as the unprocessed color coordinates $u_k$ that can be inferred in various ways from the directly measurable values using the model.

\begin{figure*}[h!]
\centering
\includegraphics[width=8cm, left]{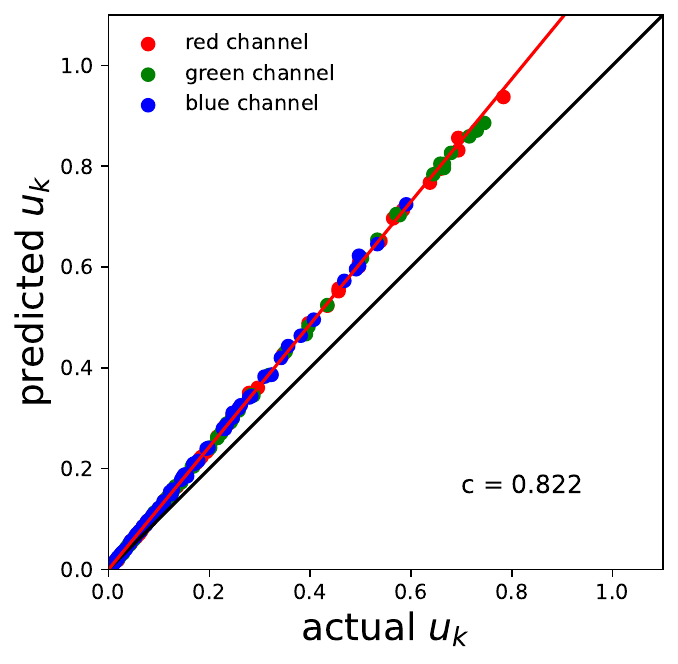}
\caption{Model predictions of unprocessed values $u_k$ from a Lambertian surface with no tonemapping, plotted against the unprocessed values inferred from actual post-processed values as $u_k = s(v_k)$. The solid black line shows $y = x$, and the solid red line is the least-squares regression line, constrained to pass through the origin.}
\label{figure:model_test_1}
\end{figure*}

As a first test of the HDRP model, Figure \ref{figure:model_test_1} compares the unprocessed color coordinates $u_k = s(v_k)$ calculated from the captured framebuffer values $v_k$, to the values of $u_k$ predicted by equation (\ref{eqn:lambert2}) for a Lambertian surface without tonemapping. Here we do not incorporate the constant $c$ in (\ref{eqn:lambert2}). The predicted and actual values are highly correlated, but they differ by a scale factor. The solid red line is the least-squares regression line, and $c$ can be estimated as the inverse of the regression slope. The value $c = 0.822$ reported earlier is based on 5000 samples, which repeated runs show to provide a stable estimate.

\begin{figure*}[h!]
\centering
\includegraphics[width=17cm, left]{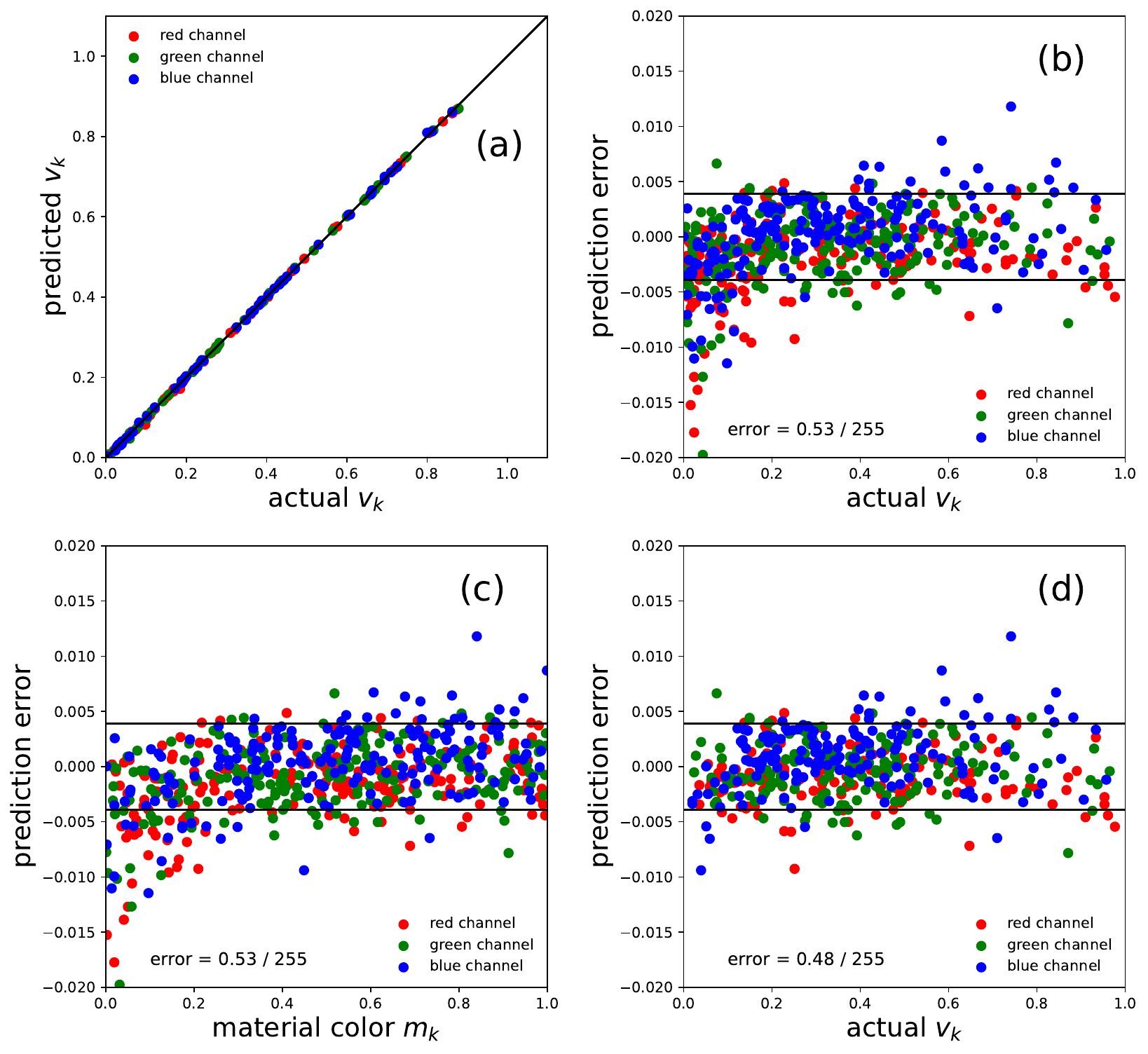}
\caption{(a) Model predictions of post-processed values $v_k$ in renderings of a Lambertian surface with no tonemapping, plotted against the actual post-processed values. The solid line shows $y = x$. (b) Prediction error for post-processed values $v_k$, i.e., predicted minus actual value. The solid lines show $y = \pm \: 1/255$. The error value is the median absolute error, reported as a multiple of 1/255. (c) Prediction error as a function of material color $m_k$. (d) Repeat of panel (b), without points with material color coordinates $m_k < 0.2$.}
\label{figure:model_test_2_L1_T0}
\end{figure*}

Figure \ref{figure:model_test_2_L1_T0} shows further analyses of data from a Lambertian surface without tonemapping. Panel (a) plots the post-processed color coordinates $v_k$ predicted by equations (\ref{eqn:lambert2}), (\ref{eqn:unlit}), and (\ref{eqn:post1}), now incorporating the scale factor $c = 0.822$, against the actual post-processed values. Overall, the model is quite accurate. Panel (b) shows the prediction error, i.e., predicted minus actual values of $v_k$, as a function of the actual values $v_k$. The horizontal lines show $y = \pm \: 1/255$, to indicate the scale of error we would expect from rounding $v_k$ to eight-bit precision. Errors are mostly within this range, and almost always within a few multiples of this range. The median absolute prediction error, expressed as a multiple of 1/255, is 0.53/255. The source of this residual error is unclear. One evident bias is that the predictions tend to be too low at low values of post-processed color coordinates $v_k$. Panel (c) shows that this bias is largely a function of the material color coordinates $m_k$: at low values of $m_k$, the predicted $v_k$ tends to be too low. Fortunately, the largest negative biases occur for material colors in the range $m_k < 0.2$, which correspond to albedos of $s(0.2) = 0.033$ and lower. It is rare for everyday materials to have albedos less than 0.03, so the strongest biases in the model occur with very dark materials that are uncommon in natural scenes. Panel (d) replots the data in panel (b), without points where $m_k < 0.2$, and shows that without these points the median absolute error is reduced to 0.48/255. Overall, these errors are small, but they could be limiting factors in experiments where observers make fine perceptual discriminations.

\begin{figure*}[h!]
\centering
\includegraphics[width=17cm, left]{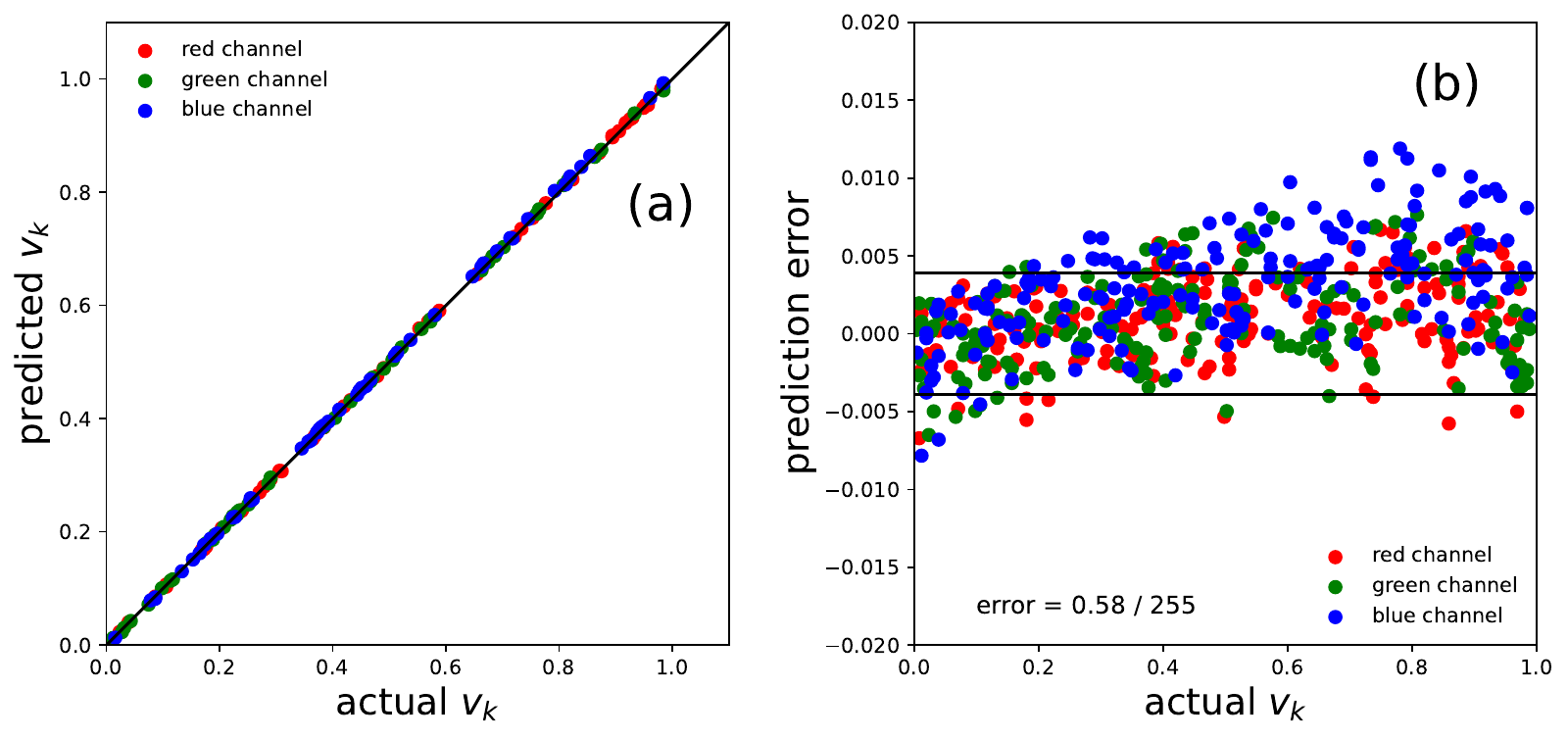}
\caption{(a) Model predictions of post-processed values $v_k$ in renderings of an unlit material type with no tonemapping, plotted against the actual post-processed values. (b) Prediction error for post-processed values $v_k$.}
\label{figure:model_test_2_L0_T0}
\end{figure*}

Figure \ref{figure:model_test_2_L0_T0} shows results of the same tests with an unlit material without tonemapping. Here too, the model is overall quite accurate. Even with this simple material, though, the predictions are not perfect, and the median absolute error is 0.58/255. For an unlit material, material color $m_k$ is the only parameter that affects the rendered value $v_k$, so any biases are again a function of $m_k$.

\clearpage

%
%

\subsection{Tests with tonemapping}

As explained in Section \ref{section:tonemapping}, tonemapping relies on a set of knot points to define the tonemapping function $\mathbf{f}$ that Unity applies to the rendered image. In order to use this feature for gamma correction, we need to know the numeric values of the knot points, which do not seem to be provided in the HDRP documentation. In this section I describe two methods of estimating the knot points, and then I test the HDRP model's predictions based on these estimates. First, in a \textit{delta function} approach, I run a series of tests where I set the tonemapping output at each individual knot point to 1.0, and the output at all other knot points to 0.0, and infer the value of the knot point from the rendered image. Second, in an \textit{optimization} approach, I treat the knot points as adjustable parameters in an optimization problem, and use numerical search routines to find the knot point values that maximize the accuracy of the HDRP model's predictions.

More precisely, as outlined in Section \ref{section:tonemapping}, tonemapping relies on a set of values $u^*_i$ that are used to generate knot points $\mathbf{u}_{ijk} = (u^*_i, u^*_j, u^*_k)$. Tonemapping maps these knot points to corresponding values $\mathbf{t}_{ijk} = (t^r_{ijk}, t^g_{ijk}, t^b_{ijk})$ that are given in a cube file. Color triplets $\mathbf{u}$ between the knot points are mapped to values $\mathbf{f}(\mathbf{u})$ by interpolation. In order to create a cube file that defines a specific tonemapping function, we need to know $u^*_i$.

\begin{figure*}[h!]
\centering
\includegraphics[width=17cm, left]{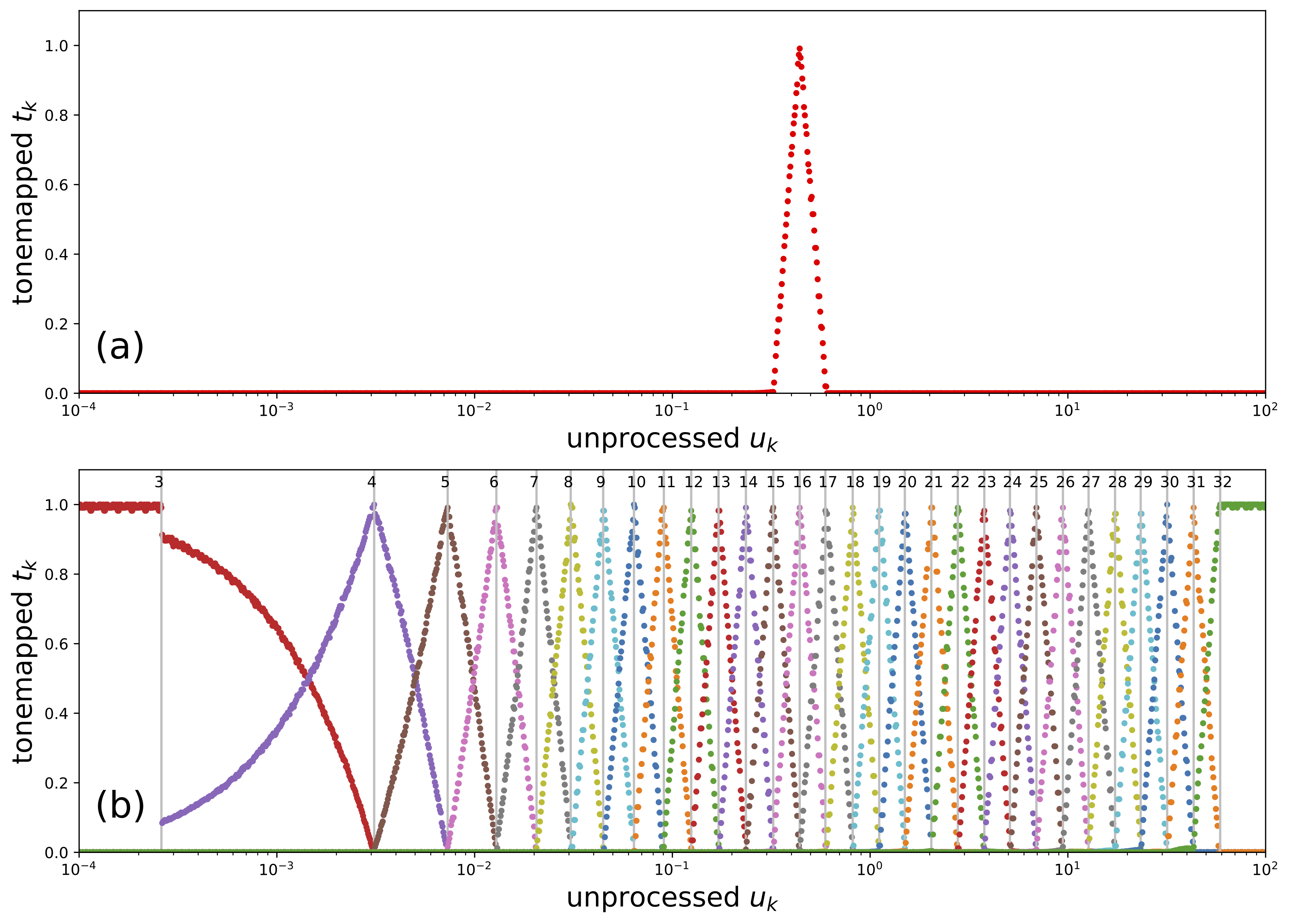}
\caption{(a) Tonemapped values $t_k$ resulting from a wide range of unprocessed values $u_k$, using a cube file that maps knot point coordinate $u^*_{16}$ to one, and all other knot points to zero. (b) The same analysis repeated 32 times, using cube files that each map a single knot point coordinate $u^*_m$ to one, and all others to zero. Results from different cube files are shown in different colors, and the peak of each curve is labelled with the number $m$ of the knot point that is mapped to one.}
\label{figure:delta}
\end{figure*}

Figure \ref{figure:delta}(a) illustrates the delta function approach to measuring $u^*_i$. Panel (a) shows the tonemapped value $t_k$, as a function of the unprocessed value $u_k$, with tonemapping carried out according to a cube file that maps knot points $\mathbf{u}_{ijk} = (u^*_i, u^*_j, u^*_k)$ to $\mathbf{t}_{ijk} = (\delta_{i,16}, \delta_{j,16}, \delta_{k,16})$, where $\delta_{ij} = 1$ if $i = j$ and 0 otherwise. That is, in each color channel the sixteenth knot point value is mapped to one, and all others are mapped to zero. With this configuration, the output of tonemapping peaks for an input value of $u_k = 0.44$, which we can take as an estimate of the knot point component $u^*_{16}$.

Figure \ref{figure:delta}(b) shows this analysis repeated with 32 cube files delta\_$m$.cube, where $m$ ranges from 1 to 32. File delta\_$m$.cube maps knot points $\mathbf{u}_{ijk} = (u^*_i, u^*_j, u^*_k)$ to $\mathbf{t}_{ijk} = (\delta_{im}, \delta_{jm}, \delta_{km})$. The peak of each dataset is labelled with the number 1-32 of the cube file that was used to create it. In each case, we can take the input value that produces the largest output as an estimate of $u^*_m$.

This analysis provides much useful information about tonemapping. First, points $u^*_1$ and $u^*_2$ do not appear to play a role. For cube files delta\_01.cube and delta\_02.cube, all input values are mapped to zero, which is why there are no peaks with labels 1 and 2 in Figure \ref{figure:delta}. Second, the knot point coordinates $u^*_i$ are approximately (though only approximately) logarithmically spaced between $u^*_3 = 0.00026$ and $u^*_{32} = 59$. Third, all input values below $u^*_3$ and above $u^*_{32}$ are mapped to a constant value. Fourth, this analysis gives estimates of $u^*_i$ for $i$ from 3 to 32, which we can use to create cube files for specific tonemapping functions. Table \ref{table:knots}(a) lists these estimates of $u^*_i$. Fifth, interpolation between neighboring points is linear, as indicated by the triangular responses to delta functions. Sixth, there is an anomaly in interpolation between points $u^*_3$ and $u^*_4$, where instead of smooth interpolation there is a discontinuity near $u^*_3$.

\begin{table}[t]
\caption{Estimates of knot points coordinates $u^*_i$ used by cube files, for $i$ from 3 to 32}
\begin{tabular}{ c c c c c c c c c c }
\hline
\hline
\multicolumn{10}{l}{(a) Estimates from delta functions} \\
\hline
0.0002606 & 0.003104 & 0.007305 & 0.01288 & 0.02056 & 0.03061 & 0.04468 & 0.06393 & 0.09056 & 0.1245 \\
0.1711 & 0.2354 & 0.3236 & 0.4406 & 0.5938 & 0.8165 & 1.111 & 1.498 & 2.039 & 2.776 \\
3.780 & 5.094 & 6.935 & 9.441 & 12.72 & 17.32 & 23.35 & 31.78 & 43.27 & 58.90 \\
\hline
\multicolumn{10}{l}{(b) Estimates from optimizing model predictions} \\
\hline
$1.657 \times 10^{-9}$ & 0.002830 & 0.007137 & 0.01269 & 0.02051 & 0.03086 & 0.04479 & 0.06444 & 0.08989 & 0.1252 \\
0.1726 & 0.2370 & 0.3253 & 0.4422 & 0.6039 & 0.8207 & 1.104 & 1.495 & 2.032 & 2.756 \\
3.738 & 5.083 & 6.864 & 9.347 & 12.62 & 17.18 & 23.24 & 31.48 & 42.75 & 57.66 \\
\hline
\hline
\end{tabular}
\label{table:knots}
\end{table}

The estimates of $u^*_i$ from this analysis (see Table \ref{table:knots}(a)) allow us to model the effect of tonemapping with cube files. Tonemapping maps knot points $\mathbf{u}_{ijk} = (u^*_i, u^*_j, u^*_k)$ to values $\mathbf{t}_{ijk}$ specified in a cube file, and maps other values of $\mathbf{u}$ by interpolation. The Supporting Information defines a Python class, TonemapCube, that loads a user-supplied cube file and implements this model of the tonemapping function $\mathbf{f}$. This makes it possible to use equation (\ref{eqn:post1}) to calculate predictions for post-processed color coordinates $v_k$ that incorporate tonemapping.

\begin{figure*}[h!]
\centering
\includegraphics[width=17cm, left]{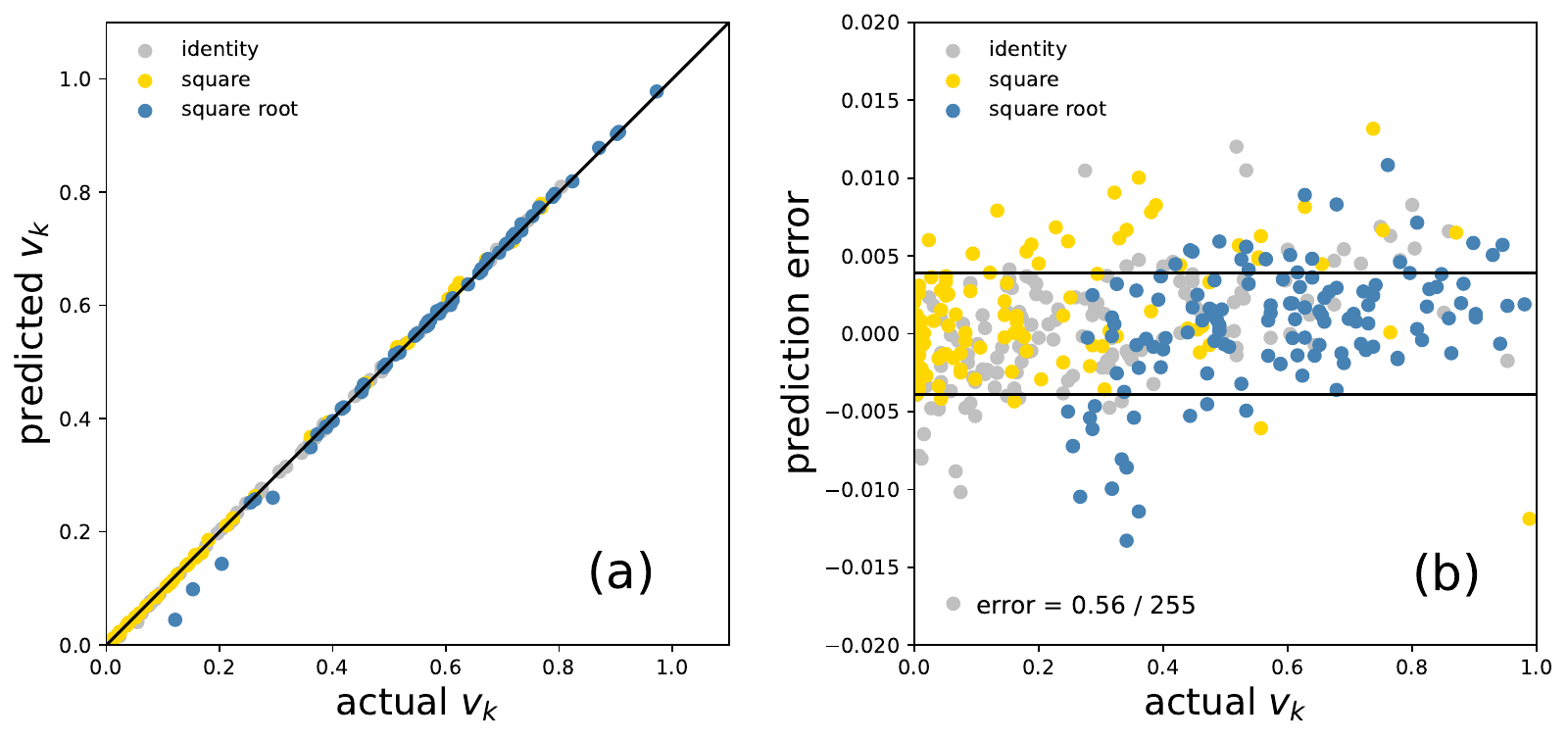}
\caption{(a) Model predictions of post-processed values $v_k$ in renderings of a Lambertian surface with three kinds of tonemapping (identity, square, and square root functions), plotted against the actual post-processed values. Tonemapping predictions were made using knot point coordinates $u^*_i$ estimated by tonemapping delta functions. (b) Prediction error for post-processed values $v_k$.}
\label{figure:model_test_L1_T1_delta}
\end{figure*}

Figure \ref{figure:model_test_L1_T1_delta} shows results of tests like those reported in Section \ref{section:tests_without_tonemapping}, with the HDRP model extended to include tonemapping. This test combines data from random, rendered scenes, generated by the render\_random project described above, with three different tonemapping functions: the tonemapped value $t_k$ is identical to, the square root of, or the square of the unprocessed value $u_k$. This gives an indication of how the model performs across a range of tonemapping functions. The model's predictions are broadly accurate for all three tonemapping nonlinearities, but show somewhat higher errors than without tonemapping. This suggests that the estimates of $u^*_i$ can be improved. Just as rendered values $v_k$ are subject to small random variations and biases (e.g., see Figure \ref{figure:model_test_2_L1_T0}), the tonemapped values $t_k$ that we used to estimate the knot point coordinates $u^*_i$ have small departures from their idealized values as well. As a result, the effects of cube files that implement delta functions reveal useful information about tonemapping, as we have seen, but they do not allow us to estimate $u^*_i$ with arbitrary precision.

An alternative, more direct approach to estimating the knot point coordinates $u^*_i$ is to search for values of $u^*_i$ that optimize the model's predictions. For this approach, I used data that consisted of random scene parameters and rendered values, generated by render\_random, with tonemapping set to identity, square root, and square functions. Samples with $m_k < 0.2$ were eliminated from the dataset, since tests without tonemapping had shown the HDRP model to be biased for these values, and I did not want the estimates of $u^*_i$ to take on biases in order to compensate for the model's biases at low $m_k$.  Starting from initial estimates of $u^*_i$ given by the delta function analysis, a nonlinear optimization routine searched for the values of $u^*_i$ that minimized the sum-of-squares error in the HDRP model's predictions. Thus this was an optimization problem where the optimized parameters were the values $u^*_i$ that are used to construct knot points, and the error to be minimized was the HDRP model's prediction error for rendered values.

\begin{figure*}[h!]
\centering
 \includegraphics[width=17cm, left]{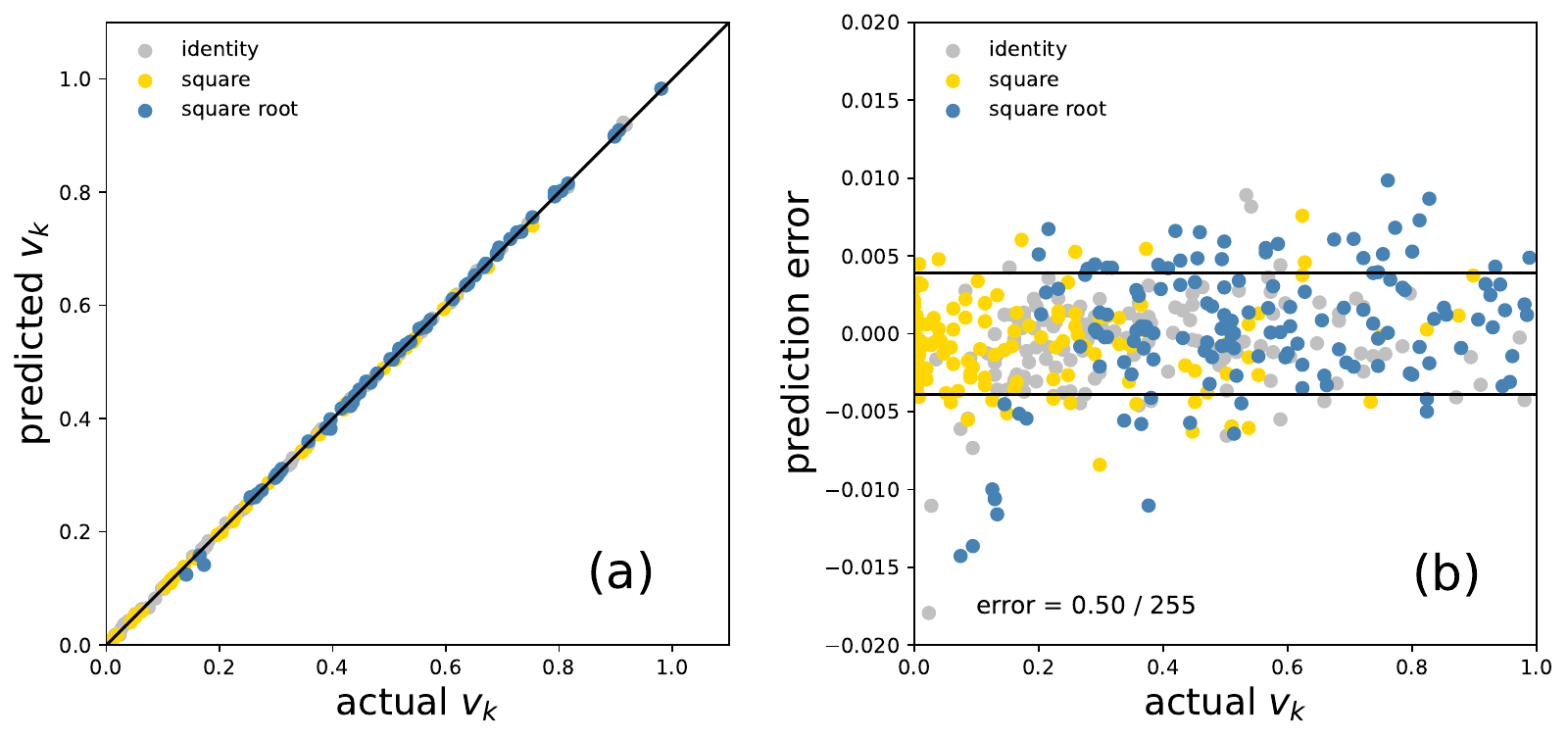}
\caption{(a) Model predictions of post-processed values $v_k$ in renderings of a Lambertian surface with tonemapping, plotted against the actual post-processed values. Tonemapping predictions were made using knot point coordinates $u^*_i$ estimated by optimizing model predictions. (b) Prediction error for post-processed values $v_k$.}
\label{figure:model_test_L1_T1_optimize}
\end{figure*}

Table \ref{table:knots}(b) reports the values of knot points estimated this way. These estimates were similar to those from the method of delta functions: the median percentage difference between knot point estimates from the two methods was just 0.8\%, and except for the smallest knot point, none of the estimates differed by more than a few percent. Nevertheless, the optimization method did provide slightly better estimates. Figure \ref{figure:model_test_L1_T1_optimize} shows the predictions and prediction error when the HDRP model with these new knot point estimates was applied to new random samples from the same dataset, i.e., samples not used to estimate $u^*_i$, in order to avoid overfitting. The model was about as accurate as the model without tonemapping, with a median absolute error of 0.50/255, compared to 0.56/255 for the delta function method. As was found for scenes without tonemapping, the largest errors occured with low material color coordinates, $m_k < 0.2$, and when these samples were eliminated, most of the larger errors disappeared, and the median absolute error was 0.49/255 (figure not shown).

\begin{figure*}[h!]
\centering
 \includegraphics[width=17cm, left]{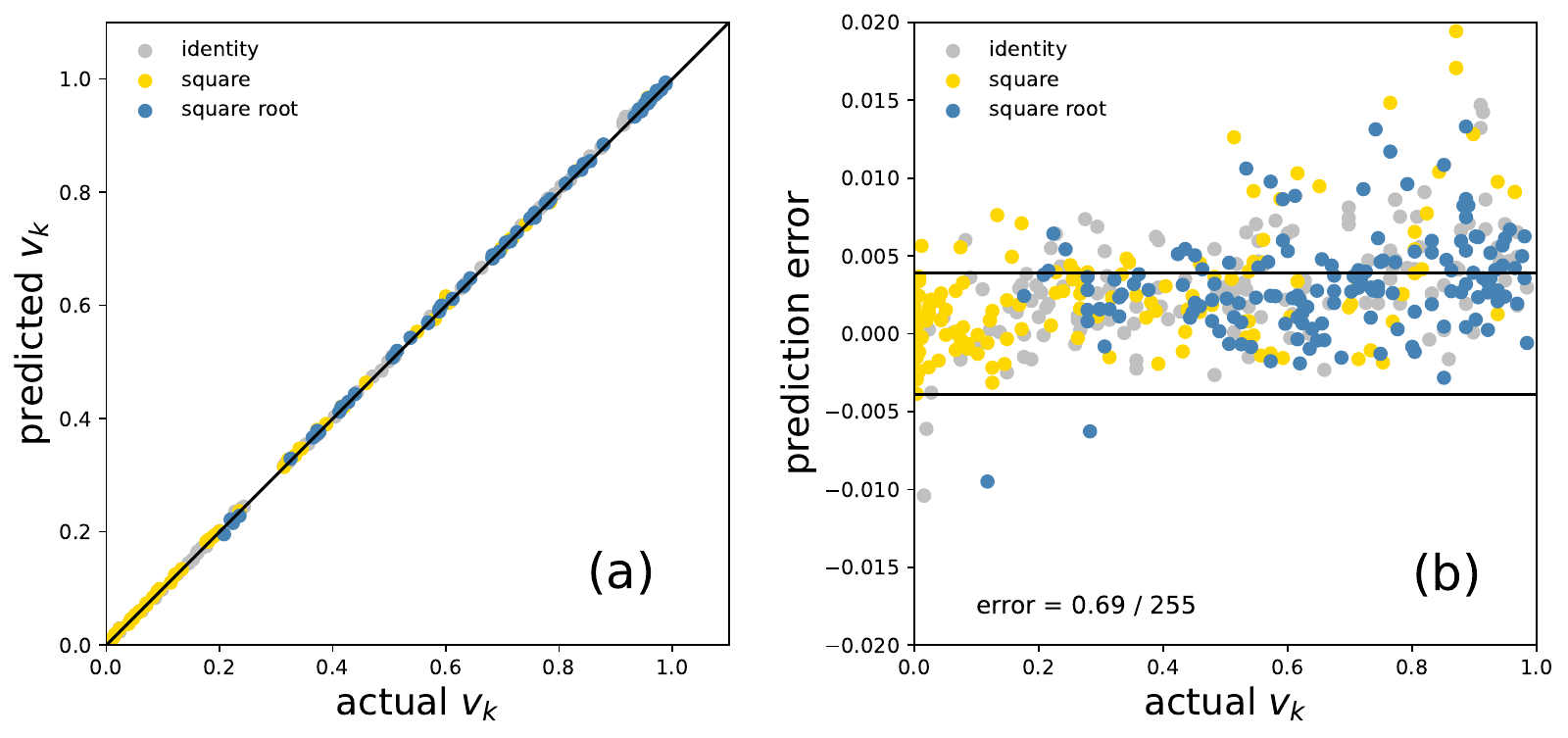}
\caption{(a) Model predictions of post-processed values $v_k$ in renderings of an unlit material with tonemapping, plotted against the actual post-processed values. Tonemapping predictions were made using knot point coordinates $u^*_i$ estimated by optimizing model predictions. (b) Prediction error for post-processed values $v_k$.}
\label{figure:model_test_L0_T1_optimize}
\end{figure*}

Figure \ref{figure:model_test_L0_T1_optimize} shows results of the same test with the same optimized knot points and an unlit material. The predictions are generally accurate, though with a somewhat higher median absolute error of 0.69/255.

\clearpage

%
%

\subsection{Tests of gamma correction} \label{section:tests_gamma}

The Supporting Information includes a Unity project (caldemo) that illustrates how to use the methods described here for gamma correction in an experiment. Unity has many configuration options, so I recommend integrating routines for characterization measurements into the experiment itself, rather having them in a separate project. This makes it less likely that there are important configuration differences between the routines where characterization measurements are made, and the experiment where they are used.

The caldemo project is a simple orientation discrimination experiment that includes characterization code. A checkbox in the Unity editor flags whether achromatic or chromatic characterization should be used. At any point during the experiment, the user can press a key to switch the program into characterization mode, at which point it shows a series of achromatic or chromatic test stimuli, for which the user can measure the luminance or XYZ color coordinates. Separate Python scripts use these measurements to generate a cube file for tonemapping, which can then be loaded into the project. When the experiment is run again and the characterization measurements are made again, luminance or XYZ coordinates are proportional to unprocessed values $u_k$. The Supporting Information includes instructions on how to carry out these tasks.

\begin{figure*}[h!]
\centering
\includegraphics[width=17cm, left]{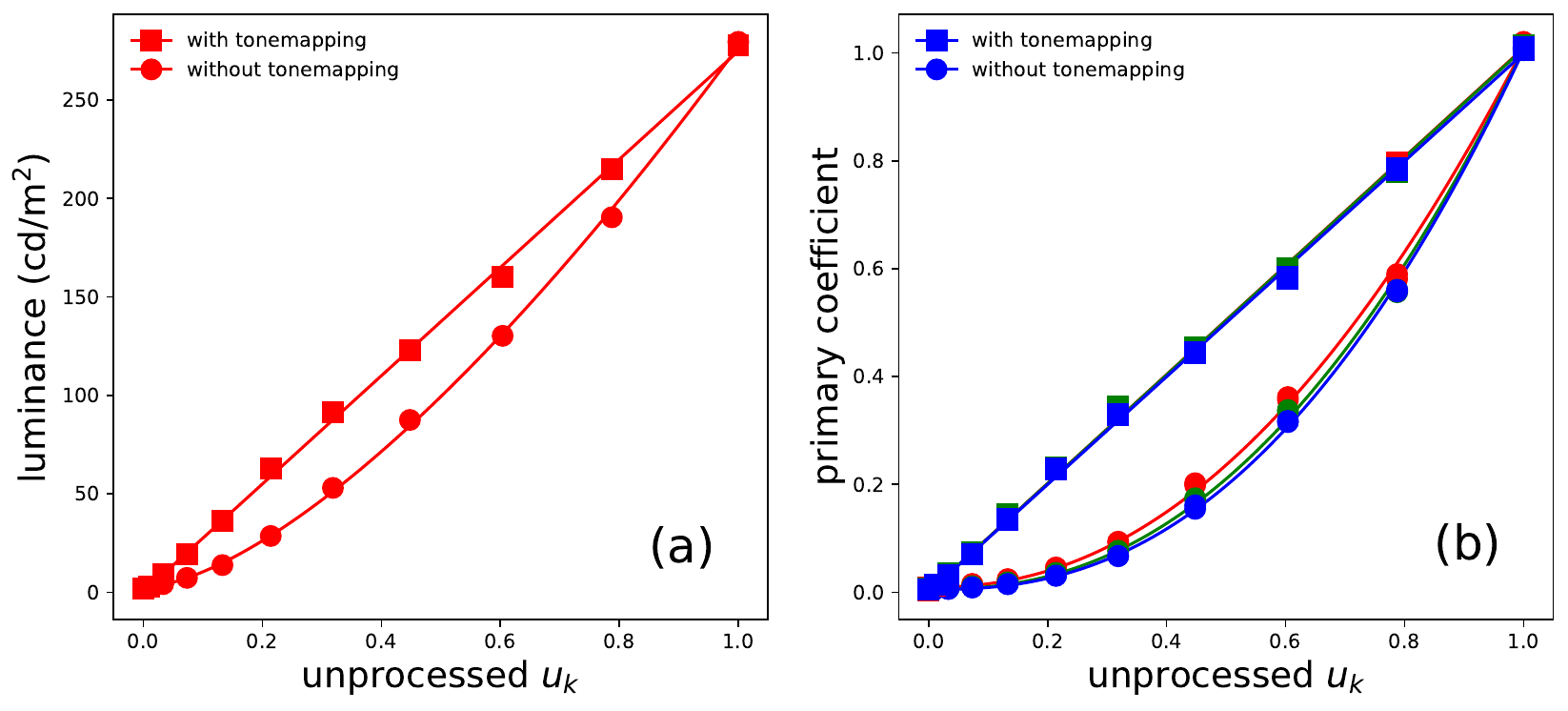}
\caption{Display characterization measurements. The curves through measurements without tonemapping are fits of a gamma function. The lines through measurements with tonemapping are fits of a straight line constrained to pass through the origin. (a) Luminance characterization.  (b) Color characterization, with each plot color representing the corresponding color channel (red, green, or blue), and the $y$-axis representing the primary coefficients in equation (\ref{eqn:chromatic_model2}).}
\label{figure:characterization}
\end{figure*}

Figure \ref{figure:characterization} shows results for these tests of achromatic and chromatic characterization. I found that several monitors had gamma functions close to the sRGB standard, and as a result the nonlinearities in the HDRP produced luminance values and XYZ coordinates that were fairly close to a linear function of unprocessed color coordinates $u_k$, even without using tonemapping for gamma correction. To show that the methods described here linearize the response even when there are significant departures from the sRGB standard, I used MATLAB and the Psychtoolbox \citep{kleiner2007} to set a monitor's lookup table so that the response was clearly nonlinear (Figure \ref{figure:characterization}). Luminance measurements were made using a Minolta LS-110 photometer, from an LCD monitor (Samsung, model 24T350) driven by an NVIDIA GeForce 2080 graphics card on a PC running Windows 11.  Figure \ref{figure:characterization}(a) shows the nonlinear response in luminance characterization measurements without tonemapping, as well as the linearized response when tonemapping was applied using a cube file generated by code provided in the Supporting Information. Figure \ref{figure:characterization}(b) shows results of a corresponding test for chromatic gamma correction, using measurements made on the same monitor using a Photo Research PR-655 spectroradiometer. In both cases, tonemapping approximately linearized the relationship between the rendered values $u_k$ and the physical stimulus. There were some small departures from linearity, which may be due to the inability of a gamma function, which was used to create the cube file, to fit the empirical measurements exactly; a more flexible function such as a spline may give slightly better results. I obtained similar results when repeating these two tests with an Oculus Pro head-mounted display, driven by an NVIDIA GeForce RTX 3090 graphics card, with Unity configured for VR as described in detail in the Supporting Information.

For this method of gamma correction to work, the display must meet the assumptions of the rendering model. I ran the same tests on an iMac (24" display, Apple M1 chip, 2021, macOS 15.3.2), and found that gamma correction failed. The reason was that the post-processed values $\mathbf{v}$ rendered by Unity were not the final framebuffer values, as revealed for instance by macOS's Digital Color Meter. That is, the operating system applied a transformation to Unity's final rendered values before displaying them. The nature of this transformation was unclear, but it was affected by the ICC color profile chosen for the display in the system preferences. As a result, each framebuffer color channel depended on all three rendered color channels $\mathbf{v}$, and since the framebuffer values were nonlinearly related to the physical stimulus, the assumption of channel additivity was violated. Modern operating systems include color management methods whose effects are not always easy to determine, and so it is important to test the success or failure of gamma correction methods.

%
%

\section{Discussion}

Game engines such as Unity and Unreal are complex programming environments, and one lesson from the present work is that it is important to have a well-tested model of color and other stimulus features that we want to control in experiments. Consider some of the idiosyncratic features that are documented in equation (\ref{eqn:lambert2}) for Unity's treatment of Lambertian materials: the nonlinearity $s$ is applied to material color coordinates $m_k$ and directional light intensity $d_k$, but not to ambient light intensity $a_k$; the effects of the parameters for directional and ambient light intensity, $i_d$ and $i_a$, differ by a factor of $\pi$; and the rendered values depend on a constant $c = 0.822$. Furthermore, rendered values $u_k$ are passed through additional nonlinearities, as in equation (\ref{eqn:post1}). A model of these processes is necessary in order to reliably control stimuli, and without such a model, errors are likely. The tests reported here also show that the rendered values have a random component, or at least a random-seeming component that is not predicted by a simple Lambertian model, and this may be an important factor in experiments on fine perceptual discriminations.

I have presented a model of the HDRP in a simple configuration, namely a Lambertian surface under directional and ambient lighting. One advantage of game engines is that they allow us to carry out experiments in complex, somewhat realistic, rendered scenes, so it will be important to extend this work to examine more advanced features of the HDRP, such as light probe illumination and more realistic material models. The Unity projects and analysis code provided as Supporting Information provide a starting point for such work.

The HDRP model that I have described is complementary to standard gamma characterization methods, rather than a replacement. Common gamma correction models, such as the gain-offset-gamma (GOG) model \citep{berns1996} and its variant described by \cite{brainard2002}, describe a display device's mapping from rendered framebuffer values to displayed stimuli in physical units. In the nomenclature of the HDRP model, they describe the mapping from post-processed color coordinates $\mathbf{v}$ to diplayed luminance $L$ or color coordinates $\mathbf{x}$. In the approach to gamma correction described above, we choose the tonemapping function $\mathbf{f}$ to compensate for this mapping and linearize the relationship between rendered values $\mathbf{u}$ and displayed luminance or color. In order to do this, we need to characterize the mapping from $\mathbf{v}$ to the physical stimulus, and fitting a model such as GOG to a limited number of display measurements is a useful way of doing this. This is the approach I took in the validation tests, reported above, of the HDRP model with tonemapping, and it is illustrated in the code in the Supporting Information. This combination of the HDRP model, which describes several nonlinearities in the Unity render pipeline, and models such as GOG that have been validated as descriptions of display nonlinearities, provides an integrated approach to modelling and generating stimuli in Unity.

Table \ref{table:variables} lists the range of possible values for variables in the rendering model. In fact, some variables can be assigned values outside these ranges, e.g., in a C\# script, a Lambertian material can be assigned color coordinates $m_k$ greater than one, which implies an an albedo greater than one. Some other variables can be assigned out-of-range values as well. However, I have not tested the model thoroughly for such values, and I do not recommend using them without further testing.

One challenge for characterizing head-mounted displays is simply measuring displayed luminances and colors. This problem is largely outside the scope of this paper, but the methods presented here depend on it. The ideal solution is to use specialized equipment that has been designed for head-mounted displays, but such equipment is currently quite costly. Many labs have spot photometers, colorimeters, or spectrophotometers, but the optics of these devices are typically not matched to head-mounted displays. \cite{murray2022} gave some evidence that a Minolta LS-110 spot photometer provides measurements that are proportional to the luminance displayed in an Oculus Rift S VR headset, but they also showed that the measurements integrate light in the headset's field of view over a region more than 15$^\circ$ in diameter, well beyond the photometer's nominal integration region of 0.33$^\circ$. The luminance and chromaticity of a display can vary appreciably over regions this large, which may compromise measurements. It would be useful to have studies that compare measurements of luminance and color in head-mounted displays, when made with specialized equipment and with more widely available tools such as spot photometers. Similarly, technical specifications for head-mounted displays often report their color primaries and gamma functions, for example as conforming to a standard such as sRGB, and it would be useful to know how accurate and stable these specifications are in practice for consumer devices.

I developed the model described in this paper for the most part empirically, by trial and error, with some guidance from Unity's documentation. As shown here, I have validated the model under some rendering conditions. Nevertheless, I encourage users to use the code provided in the Supporting Information to test the model in their own experiments, and in general to use the model cautiously. It may need to be extended for rendering conditions beyond those tested here.

\section{Acknowledgements}

I thank David Brainard and Fangfang Hong for helpful discussions. This work was supported by a Discovery Grant from the Natural Sciences and Engineering Research Council of Canada.

\begin{appendices}

\section{Configuring an HDRP project} \label{appendix:configure}

Here I describe how to create and configure a Unity project where the models described in this paper hold. I also explain what parameters of Unity lights and materials correspond to the variables in equations (\ref{eqn:lambert2}) and (\ref{eqn:unlit}). I assume the reader is familar with basic procedures for using Unity. For an introduction to Unity, I recommend the first eight chapters of \cite{geig2022}. They cover the Built-in Render Pipeline, but most of their content applies to the HDRP as well.

In the Unity Hub, create a new project using the `High Definition 3D' template. After the project opens in the Unity Editor, the default HDRP configuration is found in the project settings (menu entry Edit / Project Settings ...). Instead of changing these settings directly, we will override them in a Volume object.

In the Hierarchy view, delete the `Sky and Fog Volume'. Create a `Volume / Global Volume'. This object controls rendering properties in the scene. In the Inspector view for this volume, next to Profile, click New.

Still in the Inspector view for Global Volume, click `Add override', and choose type Exposure. This override configures the exposure feature described in the main text, that in its default state dynamically adjusts the gain on rendered images. Click the checkbox next to Mode, and leave the mode at its default value, which is Fixed. Click `Fixed Exposure', and enter a value of 0. This is the value of $e$ in equation (\ref{eqn:lambert2}), which shows that rendered values are now divided by $2^0 = 1$. 

Add another override, and this time choose type `Post-processing / Bloom'. The bloom feature causes light scattering around bright objects. Click Quality, and set it to Low. Click Intensity, and enter a value of zero. This feature is now off.

Add a `Post-processing / Tonemapping' override. Click Mode, and leave it at its default value, which is None. Tonemapping is now disabled, i.e., $\mathbf{f}(\mathbf{u}) = \mathbf{u}$. Alternatively, set the mode to External. Drag a cube file into the Assets view. (Cube files are included in the Supporting Information.) In the Inspector view for Global Volume, in the Tonemapping override, check `Lookup Texture'. Click the small circle to the right, and in the window that appears select the cube file you put in the Assets view. The tonemapping feature is now active, and it uses the mapping specified in the cube file.

The next two overrides configure the ambient lighting. First, add a `Visual Environment' override. Click `Sky Type', and choose `Gradient Sky'. Second, add a `Sky / Gradient Sky' override. This override allows the user to set the color of the top, middle, and bottom parts of the sky. We will set all three to the same value, to create a classic ambient light source with the same color and intensity in all directions. Click Top. Click the color patch to the right, and in the window that appears, set the color coordinate mode to `RGB 0-1.0', enter the value 1 for each of the three colors (these are the values $a_k$ in equation (\ref{eqn:lambert2})), and close the color picker window. Repeat this procedure for the Middle and Bottom features of the Gradient Sky override. Then click `Intensity Mode', and set it to Multiplier. Click Multiplier, and set the value to 1. This is $i_a$ in equation (\ref{eqn:lambert2}). The ambient light is now achromatic and equally strong in all directions.

To create a directional light source, delete the Sun object in the Hierarchy view, and create a `Light / Directional Light'. In the Inspector view for this light, in the Emission panel, set `Light Appearance' to Color. In the same Emission panel, the color coordinates of the Color property are $d_k$ in equation (\ref{eqn:lambert2}), and the Intensity property is $i_d$.

There is no single Unity parameter that corresponds to the lighting direction $\mathbf{l}$ in equation (\ref{eqn:lambert2}). Instead, the lighting direction is controlled by the Rotation parameters in the Inspector view of the directional light. With Rotation parameters $X = Y = Z = 0$, directional light travels from $-z$ to $+z$, i.e., in direction (0, 0, 1). From this starting point, the Rotation parameters (which are Euler angles) rotate the light source about the corresponding axes, e.g., $X = 50$ means a 50$^\circ$ rotation about the $x$-axis. There are three points to note about these rotations. First, Unity uses a left-handed coordinate system. Second, the $z$-axis rotation is applied first, then the $x$-axis rotation, and then the $y$-axis rotation. Third, the rotations are extrinsic, meaning that they are made around fixed axes, e.g., the $z$-axis rotation, which is first, does not change the orientations of the $x$- and $y$-axes that are used for the remaining two rotations. These facts, along with the requirement that $\mathbf{l}$ in equation (\ref{eqn:lambert2}) points in the direction the light comes from, imply that
\begin{equation}
\mathbf{l} = (-\cos X \sin Y, \sin X, -\cos X \cos Y)
\end{equation}
where $X$, $Y$, and $Z$ are the Rotation parameters of the directional light source. As an alternative, the Unity project render\_random in the Supporting Information shows how to control lighting direction in a C\# script using a direction vector instead of Euler angles.
 
Next we create a Lambertian object. In the Hierarchy view, create a `3D Object / Sphere'. In the Assets view, create a Material and give it a name such as `myLambertian'. In the Inspector view for this material, near the top, set the Shader to `HDRP / ArnoldStandardSurface / ArnoldStandardSurface'. With its default settings, this material type is Lambertian. In the Inspector view for this material, click the color patch to the right of `Base Color'. The three color channel values in the window that opens are the material color coordinates $m_k$ in equation (\ref{eqn:lambert2}). Leave them at their default value of 1, and close the color picker window. Drag the material from the Assets view onto the sphere in the Hierarchy view. This assigns the material type to the sphere. View the sphere in the Inspector view, and at the bottom of the panel you should see that its material type is now myLambertian.

To create an object with an unlit material, follow the steps in the previous paragraph, but set the Shader to `HDRP / Unlit'. In the Inspector view for this material, in the `Surface Inputs' panel, the color coordinates of the Color property are $m_k$ in equation (\ref{eqn:unlit}).

In the Built-in Render Pipeline, Unity offers a choice between `Gamma' and `Linear' color spaces, and some of the nonlinearities in the model described here are applied only in Linear mode. At the time of writing, only the Linear color space is available in the HDRP.

\end{appendices}

\bibliographystyle{apacite}
\bibliography{references.bib}

\end{document}